\documentclass[%
 aip,
 jmp,%
 amsmath,amssymb,
 reprint,%
]{revtex4-1}

\usepackage{graphicx}
\usepackage{dcolumn}
\usepackage{bm}
\usepackage{amsmath}
\usepackage{xr}
\makeatletter
\newcommand*{\addFileDependency}[1]{
  \typeout{(#1)}
  \@addtofilelist{#1}
  \IfFileExists{#1}{}{\typeout{No file #1.}}
}
\makeatother

\newcommand*{\myexternaldocument}[1]{%
    \externaldocument{#1}%
    \addFileDependency{#1.tex}%
    \addFileDependency{#1.aux}%
}
\myexternaldocument{supplemental}
\draft
\begin{document}


\title{Magnetic domain wall substructures in Pt/Co/Ni/Ir multi-layers}

\author{Maxwell Li}%
 \email{mpli@andrew.cmu.edu.}
 \affiliation{Department of Materials Science \& Engineering, Carnegie Mellon University,\\ Pittsburgh, PA 15213 USA.}

\author{Anish Rai}%
 \affiliation{Department of Physics and Astronomy/MINT Center, The University of Alabama,\\ Tuscaloosa, AL 35487, USA.}
\author{Ashok Pokhrel}%
 \affiliation{Department of Physics and Astronomy/MINT Center, The University of Alabama,\\ Tuscaloosa, AL 35487, USA.}
\author{Arjun Sapkota}%
 \affiliation{Department of Physics and Astronomy/MINT Center, The University of Alabama,\\ Tuscaloosa, AL 35487, USA.}
\author{Claudia Mewes}%
 \affiliation{Department of Physics and Astronomy/MINT Center, The University of Alabama,\\ Tuscaloosa, AL 35487, USA.}
\author{Tim Mewes}%
 \affiliation{Department of Physics and Astronomy/MINT Center, The University of Alabama,\\ Tuscaloosa, AL 35487, USA.}
\author{Di Xiao}%
 \affiliation{Department of Physics, Carnegie Mellon University,\\ Pittsburgh, PA 15213 USA.}
 \affiliation{Department of Materials Science \& Engineering, Carnegie Mellon University,\\ Pittsburgh, PA 15213 USA.}
\author{Marc De Graef}%
 \affiliation{Department of Materials Science \& Engineering, Carnegie Mellon University,\\ Pittsburgh, PA 15213 USA.}
 \author{Vincent Sokalski}%
 \affiliation{Department of Materials Science \& Engineering, Carnegie Mellon University,\\ Pittsburgh, PA 15213 USA.}

\date{\today}

\begin{abstract}
We examine the substructures of magnetic domain walls (DWs) in [Pt/(Co/Ni)$_M$/Ir]$_N$ multi-layers using a combination of micromagnetic theory and Lorentz transmission electron microscopy (LTEM). Thermal stability calculations of Q=$\pm$1 substructures (2-$\pi$ vertical Bloch lines (VBLs) and DW skyrmions) were performed using a geodesic nudged elastic band (GNEB) model, which supports their metastability at room temperature. Experimental variation in strength of the interfacial Dzyaloshinskii-Moriya interaction (DMI) and film thickness reveals conditions under which these substructures are present and enables the formation of a magnetic phase diagram. Reduced thickness is found to favor Q=$\pm$1 substructures likely due to the suppression of hybrid DWs.  The results from this study provide an important framework for examining 1-D DW substructures in chiral magnetic materials.

\end{abstract}

\pacs{Valid PACS appear here}
\maketitle

\section{\label{sec:level1}Introduction}

Discovery of the Dzyaloshinskii-Moriya Interaction (DMI) in bulk magnets\cite{Dzyaloshinsky1958,Moriya1960} and magnetic thin films\cite{Thiaville2012} has launched an intense research effort into its effects on the structure of magnetic bubbles and domain walls including the formation of topological excitations like skyrmions\cite{Yu2010,Huang2012,Woo2016,Pollard2017} and anti-skyrmions\cite{Heinze2011,Nayak2017}. The combination of topological protection, which offers improvement to thermal stability, and their susceptibility to manipulation by spin-orbit torques makes these features of great interest for future spintronic applications including non-volatile memory \cite{Tomasello2014,Hanneken2015,Vakili2020} and neuromorphic computing\cite{Jiang2015,Huang2017,Li2017,Song2020}. 
 
\begin{figure}[b!]
\includegraphics{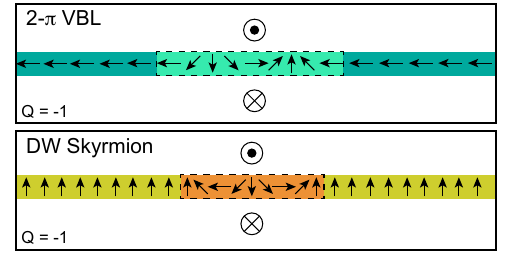}%
\caption{\small\label{Fig:schem} Representative schematics depicting the internal magnetization of 2-$\pi$ vertical Bloch line along an achiral Bloch wall and a domain wall skyrmion along a chiral N\'eel wall. Note that topological charge, $Q$, is determined by following the magnetizations along the domain wall (L to R) rather than across it.}%
 \end{figure}

While it is well established that the internal structure of a domain wall (DW) transitions from Bloch to N\'eel type with increasing interfacial DMI, less attention has been paid to its effect on the internal DW substructure. A feature known as a magnetic DW skyrmion has recently been theoretically predicted\cite{Cheng2019} and is a 360$^\circ$ rotation of the internal magnetization of a chiral N\'eel DW (Fig.~\ref{Fig:schem}). These are the post-DMI analogue of vertical Bloch lines (VBL), which are 180$^\circ$ rotations in a Bloch DW and were once considered for universal computer memory\cite{Konishi1983}. It is notable that DW Skyrmions are predicted to be about one order of magnitude smaller than VBLs. \cite{Cheng2019} Both of these substructures are confined to move within a magnetic DW and are, therefore, not subject to edge pinning (like a conventional DW) and are not able to drift in unwanted directions as with 2-D skyrmions (via the skyrmion Hall effect). In addition to their potential use in non-volatile magnetic memory storage or neuromorphic computing applications, these DW substructures have been shown to affect DW motion\cite{Yoshimura2016} and the formation of skyrmions via stripe pinching\cite{Garlow2020} and nucleation\cite{Lepadatu2020,Je2021}.

Both DW skyrmions and VBLs can be described by their topological charge as calculated from $4\pi\,Q=\int\mathrm{d}x\mathrm{d}y\,\mathbf{m} \cdot \left( \partial_x \mathbf{m} \times \partial_y \mathbf{m} \right)$, where $\mathbf{m}$ is the unit magnetization vector\cite{nagaosa2013}. A DW skyrmion has a topological charge of $Q=\pm1$ whereas a single VBL has a charge of $Q=\pm1/2$. In the case of a 2-$\pi$ VBL (i.e., a 360$^\circ$ rotation within a Bloch DW), the topological charge is equivalent to that of a DW skyrmion.  Herein, DW skyrmions and 2-$\pi$ VBLs are collectively referred to as Q=$\pm$1 substructures.

 \begin{figure*}[t!]
 \includegraphics[width=1.0\textwidth]{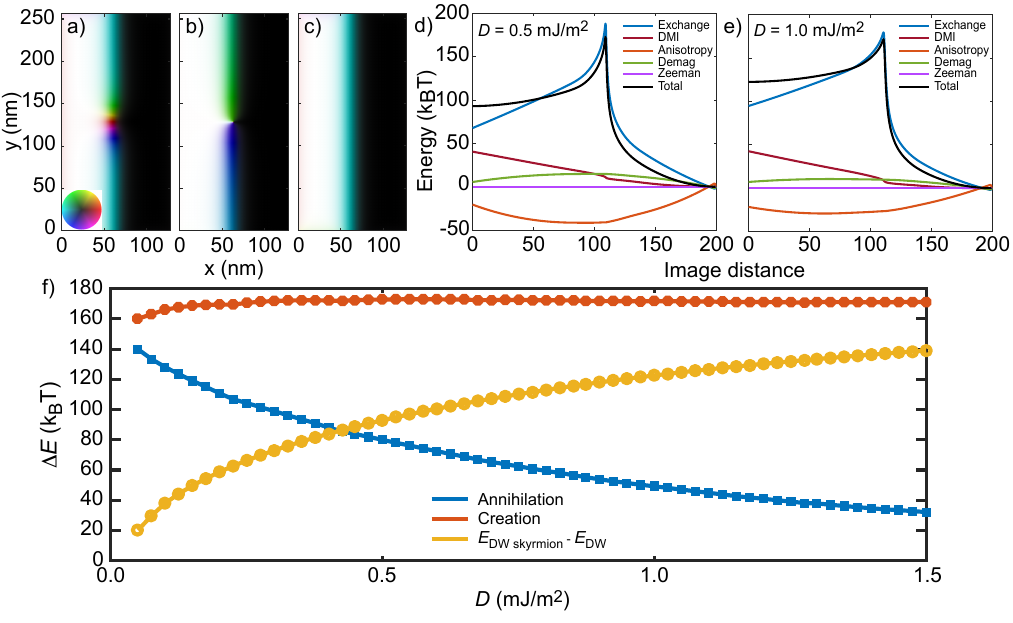}%
\caption{\small\label{Fig:NEB} Micromagnetic outputs of a Dzyaloshinskii domain wall a) with a domain wall skyrmion (image distance = 0), b) upon annihilation (image distance = 107), and c) without one (image distance = 200). d,e) Energy contributions as a function of the image distance between the domain wall skyrmion solution and the domain wall without a domain wall skyrmion.  f) Results from Nudged Elastic Band calculations of creation and annihilation barriers for a domain wall skyrmion as a function of DMI strength.}%
 \end{figure*}

Here, we examine both VBLs and DW skyrmions in perpendicularly magnetized thin films. Initial thermal stability calculations were performed via micromagnetic simulations to evaluate the metastability of VBLs and DW skyrmions. The presence of such substructures is expected to be strongly dependent on DMI strength as well as the thickness of the thin films, which is described in more detail in a subsequent section. As such, we leverage a highly tunable asymmetric multi-layer system based on (Pt/[Co/Ni]$_{M}$/Ir)$_{N}$, where a reduction in `$M$' leads to a greater interfacial DMI from the Pt/Co and Ni/Ir interfaces and `$N$' primarily modulates the total film thickness, to identify the optimal conditions where VBLs and DW skyrmions may exist\cite{Chen2013,MoreauLuchaire2016,Li2019,Li2020}. These samples were imaged using Lorentz transmission electron microscopy (LTEM) which offers high resolution imaging of magnetic features in thin films. The results of this systematic study of $M$ and $N$ iterations were used to formulate a magnetic phase diagram describing substructures observed as a function of DMI strength and sample thickness.

\section{Thermal Stability Simulations}
 
\subsection{Geodesic Nudged Elastic Band Model}

In order to utilize these DW substructures for possible spintronic applications, an understanding of their stability is necessary as thermal fluctuations can lead to their annihilation. Prior treatment of DW substructures considered the energy of the VBLs or DW Skyrmions and their ground state configurations, but did not consider the energy barrier associated with their formation or annihilation\cite{Cheng2019}. Here, we have employed a geodesic nudged elastic band (GNEB) method\cite{Bessarab2015} in combination with a climbing image method\cite{Henkelman2000} implemented in the micromagnetic code M$^3$, a MATLAB code based on finite-differences\cite{M3}. The GNEB we use builds on the nudged elastic band model (NEB) but takes the constraint into account that the saturation magnetization of each cell in the simulation volume remains constant. For N magnetic moments this method results in an unconstrained optimization within a 2N dimensional Riemannian manifold, as is discussed in detail in reference \cite{Bessarab2015}. For the evaluation of the geodesic distance between two images we use Vincenty's formula \cite{Vincenty1975}. In order to converge the images to the nearest minimum energy path we use a steepest descent method \cite{Exl2014} with a Barzilai-Borwein step length selection method \cite{Barzilai1988}. To determine their stability one has to find the activation barrier which separates the skyrmion state from lower energy states. In the case of conventional skyrmions this would be the skyrmion state and the simple ferromagnetic state. For the case of a DW skyrmion the corresponding lower energy state is a skyrmion-free domain wall. Since thermally activated magnetic transitions are rare events, dynamical simulations using a stochastic Landau-Lifshitz-Gilbert equation are not practical. Therefore the GNEB method is used to find the minimum energy path for the transition; this has been successfully applied to study the annihilation of conventional magnetic skyrmions\cite{Lobanov2016, Sampaio2013}. 

\subsection{Simulated Results and Discussion}

To stabilize the DW substructure we consider an ultra-thin ferromagnetic film ($2$ nm) with an interfacial DMI interaction and a uniaxial perpendicular anisotropy\cite{Cheng2019}. The total volume simulated was $128\times256\times2$ nm where the cell size was $0.5\times0.5\times2$ nm. Magnetic parameters of $M_\text{S}=600$ kA/m, $K_\text{eff}=2\times10^5$ J/m$^3$, $A=1.6\times10^{-11}$ J/m, and $D=0.2-1.5$ mJ/m$^2$ were used, which are similar to those measured in the samples examined in this work. The symmetric exchange and the dipole-dipole interaction are included in the micromagnetic simulations. 

To calculate the minimum energy path between those two states, 200 images were created to represent the transition path between the two fixed endpoint images. Fig.~\ref{Fig:NEB}b) shows the domain wall with a DW skyrmion, and Fig.~\ref{Fig:NEB}c) shows the domain wall after the annihilation of the substructure. The transition between these states occurs through a sharp narrowing of the DW skyrmion before the center spin flips direction concurrent with a change in the topological charge to $Q=0$. 

The energy barrier to annihilation is determined by the difference in total energy between the first image and the maximum energy encountered during the annihilation process. This was done for a range of DMI strengths whereby the barrier to annihilation is observed to decrease with increased $D$ and the barrier to creation increases initially but levels off with large $D$ (Fig. \ref{Fig:NEB}f)). We note that a VBL with $D=0$ was not examined as the only pathway to annihilation for a 1-$\pi$ VBL is via the edge. However, as previously shown by Cheng, et al., a critical DMI strength of 0.125 mJ/m$^2$ is necessary to stabilize a full 360 degree winding with these parameters\cite{Cheng2019}. More notably, even for the ultrathin film considered here, the energy barrier to annihilation is $> 60k_BT$ for $D < 1.0$ mJ/m$^2$ at room temperature. This energy barrier is directly rooted in the symmetric exchange (i.e., the exchange stiffness) as with 2D skyrmions. In the thin film approximation (i.e., uniform magnetization through the thickness), this value should scale linearly with thickness. However, it is reasonable to expect that dipolar interactions in substantially thicker films could lead to the formation of hybrid DWs characterized by a magnetization that rotates through the film thickness.\cite{Suzuki1978,Legrand2018,Dovzhenko2018} This would inherently create low energy pathways for the DW skyrmion to annihilate as further discussed in the experimental section to follow.  

\section{Lorentz TEM Imaging of Domain Wall Substructures}

\begin{figure}[t!]
 \includegraphics[width=0.475\textwidth]{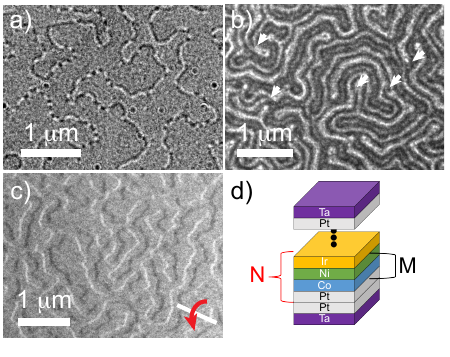}%
\caption{\small\label{Fig:PRM} Fresnel mode Lorentz TEM micrographs of symmetric a) $M=10$ and b) $M=100$ samples. Arrows in b) indicate locations of vertical Bloch lines denoted by contrast reversal from white/black to black/white (and vice versa) as seen more clearly in a) where 2-$\pi$ vertical Bloch lines are present. c) Fresnel mode Lorentz TEM micrographs of asymmetric $M=2$, $N=20$ sample in the tilted state. d) Schematic of the asymmetric Pt/Co/Ni/Ir stack whereby $M$ designates the number of Co/Ni repeats in each repetition of Pt through Ir, $N$.} 
 \end{figure}

\begin{figure*}[t!]
 \includegraphics{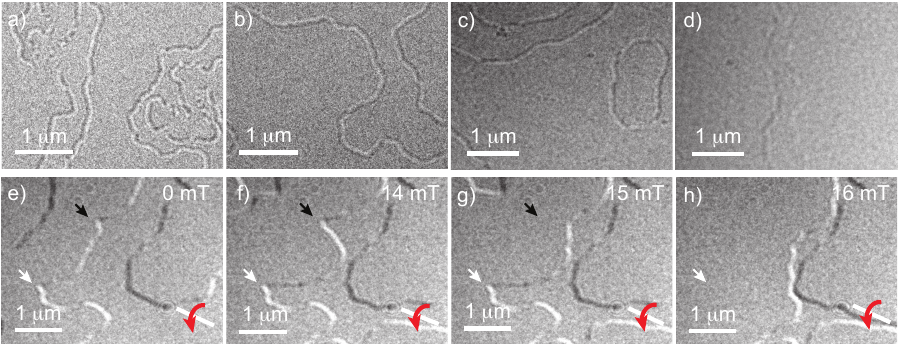}%
\caption{\small\label{Fig:lowM} Fresnel mode Lorentz TEM micrographs of $M=4-10$, $N=1$ samples where $M=$ a) 10, b) 8, c) 6, and d) 4 depict Bloch domain walls suggesting relatively small DMI strength is displayed. Contrast reversal from white/black to black/white (and vice versa) along domain walls indicate locations of vertical Bloch lines and 2-$\pi$ vertical Bloch lines. e-h) Fresnel mode Lorentz TEM micrographs of $M=3$, $N=2$ sample in the presence of increasing perpendicular magnetic field. Black and white arrows indicate locations of domain wall skyrmions that are seen to pin domain wall motion.}
 \end{figure*}

\subsection{Experimental Methods}
[Pt($0.5$~nm)/(Co($0.2$~nm)/Ni($0.6$~nm))$_M$/ Ir($0.5$~nm)]$_N$ multi-layers were prepared via rf (Ta layers) and dc (Pt, Co, Ni, Ir layers) magnetron sputtering on oxidized Si substrates for magnetic property measurements and 10 nm thick amorphous Si$_3$N$_4$ TEM membranes (Norcada) for LTEM imaging.  The working pressure was fixed at $2.5 \times 10^{-3}$ Torr of Ar. All samples had a Ta(3 nm)/Pt(3 nm) seed/adhesion layer and were capped with Ta(3 nm). Base pressure was maintained at less than $3\times 10^{-7}$ Torr. Magnetic properties were examined using alternating gradient field magnetometry (AGFM) and vibrating sample magnetometry (VSM), which confirms a strong perpendicular magnetic easy axis in all cases (Fig.~\ref{SupMHloops})\cite{Hellwig2007}. These films were imaged using LTEM using an aberration-corrected FEI Titan G2 80-300 at an accelerating voltage of 300 kV in Lorentz mode (objective lens off). LTEM employs the inherent in-plane magnetic induction of the sample to deflect the electron beam and form magnetic contrast. The resulting contrast formed can give further information about the DW character. Fresnel mode LTEM images shown here were taken with a defocus value of $-1.0$ to $-7.0$ mm depending on the thin film examined.

\subsection{Lorentz TEM Images and Magnetic Phase Diagram}

\begin{figure*}[t!]
\includegraphics{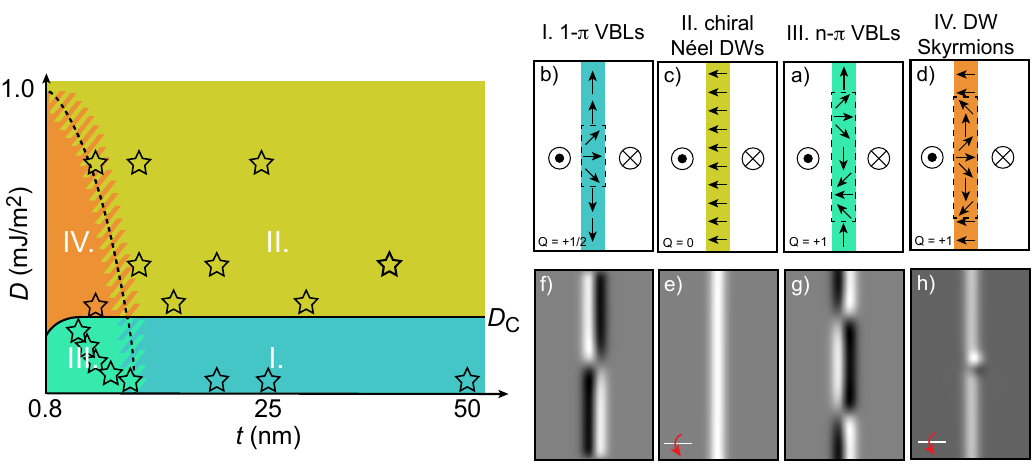}%
\caption{\small\label{Fig:phase} Qualitatively formulated phase diagram depicting conditions where the domain wall substructures depicted in a-d) can be expected to be observed with respect to DMI strength and thickness ($M\times N$). Stars on the phase diagram demark one of fifteen iterations of the Pt/Co/Ni/Ir multi-layer examined. a-d) Representative schematics and e-h) predicted Lorentz TEM micrographs of the corresponding regions are shown on the right.}%
\end{figure*}

We have previously examined the magnetic structure of this Pt/Co/Ni/Ir multi-layer system and found the formation of labyrinthine domains in relatively thick ($\geq25$nm) samples for any combination of $N$ and $M$\cite{Li2019}. It was observed that the presence of Pt and Ir sandwiching Co/Ni layers induced an appreciable interfacial DMI manifested in the form of chiral N\'eel DWs for $M=1-3$ (Fig.~\ref{Fig:PRM}). Symmetric samples did not have an appreciable DMI strength and instead displayed achiral Bloch DWs. In addition to these Bloch DWs, thick symmetric samples show sparsely distributed 1-$\pi$ VBLs and no 2-$\pi$ VBLs. This suggests an inherent instability of 2-$\pi$ VBLs, which would otherwise form when two 1-$\pi$ VBLs come into close proximity. It is worth noting, again, that such 2-$\pi$ VBLs are topologically equivalent to a DW skyrmion. It is reasonable to speculate that the relatively thick multi-layers examined here develop a non-uniform magnetization through the film thickness, which would create low energy pathways to annihilation for both 2-$\pi$ VBLs and DW skyrmions. 

Noting that neither DW Skyrmions nor 2-$\pi$ VBLs are stable in thicker multi-layer samples, we turn to thinner films where it is more reasonable to expect a uniform magnetization through the film thickness, which may support them. In thinner symmetric samples ($M=10$, $N=1$), we found a larger presence of 1-$\pi$ VBLs than in thicker ones ($M=100$, $N=1$) as seen in Fig.~\ref{Fig:PRM}.  More importantly, we note significant VBL pile-ups which  include many 2-$\pi$ VBLs. This suggests that any low-energy paths to annihilation of 2-$\pi$ VBLs associated with the larger thickness have been suppressed. Therefore, we postulate that reducing the total film thickness of similar asymmetric films while maintaining a sufficiently large DMI to support chiral N\'eel DW formation would lead to DW skyrmion formation. Having established broadly the importance of sample thickness and DMI on the stability of these excitations, we now expand on our experimental characterization of this material system through systematic variation of both these critical properties towards the development of a DW substructure phase diagram.

We began by examining thin ($N=1-2$) asymmetric films of $M=3-10$ to minimize overall film thickness. Although DMI strength increases with decreasing $M$, we observe some Bloch component to the DWs in each of these films (Fig.~\ref{Fig:lowM}a-d). Additionally, we note an overall reduction in contrast at the DW with decreasing $M$, which likely stems from both an increase in the N\'eel character of the DW as well as a reduced overall film magnetization. Even though contrast consistent with 2-$\pi$ VBL pileups are observed in each of these films, it is difficult to ascertain the true magnetic character of these substructures because the DWs are likely of mixed Bloch-N\'eel character. For $M=3$, $N=2$, exclusively chiral N\'eel DWs are observed. The Fresnel-mode contrast locally matches that predicted for DW skyrmions from micromagnetic simulations including a an asymmetric distortion of the DW itself. (Fig.~\ref{Fig:lowM}e). Moreover, we examine the response of multiple similar sites \textit{in-situ} as a perpendicular magnetic field is applied (Fig.~\ref{Fig:lowM}e-h).  It is reasonable to expect (as previously understood for VBLs)\cite{Malozemoff1972,Slonczewski1974,Krizakova2019} that any substructure should have a pinning effect on the overall DW motion. Indeed, it is seen in all cases that the defect-free portions of the wall bow around the point of interest. Eventually the wall does break past and the local dipole like contrast vanishes. We also note the absence of contrast where the DW was initially pinned confirming the absence of any larger microstructural defect (Fig.~\ref{SI_void}). These observations all point towards the existence of DW skyrmions. From the systematic examination of these different iterations, we now propose a phase diagram of DW substructures with respect to sample thickness ($M\times N$) and DMI strength (Fig.~\ref{Fig:phase}).

We divide the diagram into the four regimes described in Fig.~\ref{Fig:phase} based on our observations. Region I ($\uparrow$ DMI, $\uparrow$ thickness) contains magnetic domain walls that have a non-zero Bloch component, which include 1-$\pi$ VBLs only.  Region II ($\downarrow$ DMI, $\uparrow$ thickness) contains only chiral N\'eel DWs with no substructure.  Region III ($\downarrow$ DMI, $\downarrow$ thickness) contains magnetic domain walls that have a non-zero Bloch component characterized by highly variant DW substructures including 2-$\pi$ VBLs and VBL pileups.  Region IV ($\uparrow$ DMI, $\downarrow$ thickness) contains chiral N\'eel DWs with isolated DW skyrmions.  While the points correspond to experimentally investigated samples, the lines denote qualitative transitions between regions.  The transitions between I \& II as well as III \& IV correspond simply to the critical DMI strength required to overcome the DW anisotropy stabilizing Bloch DWs.  The transitions between I \& III as well as II \& IV correspond to the approximate thicknesses where hybrid DWs exist and create low energy paths to annihilation of DW skyrmions or 2-$\pi$ VBLs.

\section{Summary}
In summary, we modeled the thermal stability of DW substructures in Pt/Co/Ni/Ir multi-layers and experimentally characterized the parameter space (DMI and thickness) where they may exist. Thermal stability calculations revealed high energy barriers to annihilation for both DW skyrmions and VBLs, most notably $>60$ k$_B$T for $D<1.0$ mJ/m$^3$ within the approximation that magnetization is uniform through the film thickness. Lorentz TEM examination of different iterations of our quaternary system revealed four regions characterized by the DW type and substructure. Increased thickness, which is known to support the formation of hybrid DWs, inhibits stabilization of the topologically equivalent 2-$\pi$ VBLs and DW skyrmions in low and high DMI samples, respectively.  Upon reducing thickness, low energy paths to annihilate these features are suppressed and, in the high DMI case, DW skyrmions are observed. DW skyrmions are observed to locally pin the DW as expected due to their associated increase in elastic energy. Although our observations were made in Pt/Co/Ni/Ir multi-layers, which serves as a suitable test-bed, we expect that these results are not unique to this material system.\\

\section*{\label{sec:level0}Supplementary Material}
Supplementary material contains M-H hysteresis loops, calculated $M_S$ and $K_\text{eff}$, and additional Lorentz TEM images.

\begin{acknowledgments}
This research was supported by the Defense Advanced Research Projects Agency (DARPA) program on Topological Excitations in Electronics (TEE) under grant number D18AP00011. The authors also acknowledge use of the Materials Characterization Facility at Carnegie Mellon University supported by grant MCF-677785.
\end{acknowledgments}

\section*{\label{sec:level0}Data Availability Statement}
The data that supports the findings of this study are available within the article and its supplementary material.

\bibliography{aipsamp}

\end{document}


\title{Supplementary material for ``Magnetic domain wall substructures in Pt/Co/Ni/Ir multi-layers"}

\author{Maxwell Li}%
 \email{mpli@andrew.cmu.edu.}
 \affiliation{Department of Materials Science \& Engineering, Carnegie Mellon University, Pittsburgh, PA 15213 USA.}
\author{Anish Rai}%
\author{Ashok Pokhrel}%
\author{Arjun Sapkota}%
\author{Claudia Mewes}%
\author{Tim Mewes}%
 \affiliation{Department of Physics and Astronomy/MINT Center, The University of Alabama, Tuscaloosa, AL 35487, USA.}

\author{Di Xiao}%
 \affiliation{Department of Physics, Carnegie Mellon University, Pittsburgh, PA 15213 USA.}
 \affiliation{Department of Materials Science \& Engineering, Carnegie Mellon University, Pittsburgh, PA 15213 USA.}
\author{Marc De Graef}%
\author{Vincent Sokalski}%
 \affiliation{Department of Materials Science \& Engineering, Carnegie Mellon University, Pittsburgh, PA 15213 USA.}

\maketitle

\newpage
\section*{Magnetic Properties of Multi-Layers Examined}

\begin{figure}[h]
\centering
\includegraphics{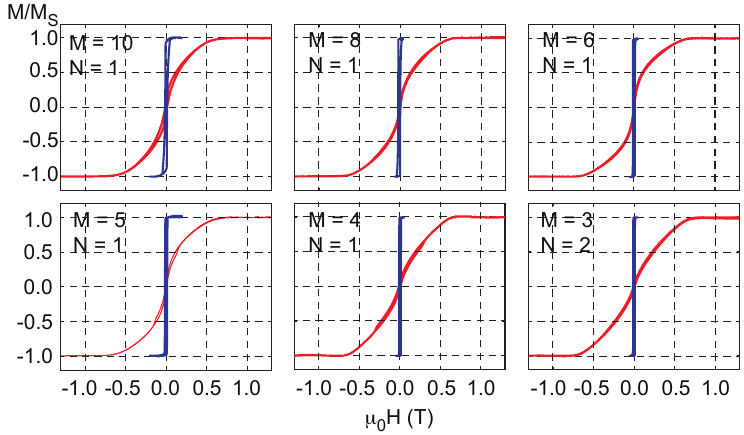}%
\caption{\label{SupMHloops} Representative parallel and perpendicular M-H loops of [Pt/(Co/Ni)$_M$/Ir]$_N$ multi-layers examined in this work.}%
\end{figure}

\noindent Representative M-H loops of [Pt/(Co/Ni)$_M$/Ir]$_N$ multi-layers examined in this work are shown in Fig. \ref{SupMHloops} which indicate strong perpendicular magnetic anisotropy in the samples examined. Relevant magnetic parameters derived from these measurements are detailed in Table \ref{tab:table1}.

\begin{table}[h]
  \caption{\label{tab:table1} Measured values of $K_\text{eff}$ and $M_\text{S}$ via VSM.}
  \begin{ruledtabular}
  \begin{tabular}{c|c||c c}
  $M$ & $N$  &   $K_{\text{eff}}$ (J/m$^3$)    &   $M_\text{S}$ (kA/m)\\
  \hline%
  10 & 1 & $3.73\times10^5$ & 1.00$\times10^3$\\
  8 & 1 & $3.66\times10^5$ & 8.10$\times10^2$\\
  6 & 1 & $4.07\times10^5$ & $7.40\times10^2$\\
  5 & 1 & $2.30\times10^5$ & $8.20\times10^2$\\
  4 & 1 & $5.23\times10^5$ & $9.03\times10^2$\\
  3 & 2 & $5.56\times10^5$ & 8.50$\times10^2$\\
 \end{tabular}
 \end{ruledtabular}
 \end{table}

\newpage
\section*{Additional Fresnel mode Lorentz TEM Images}
\begin{figure}[h]
\centering
\includegraphics{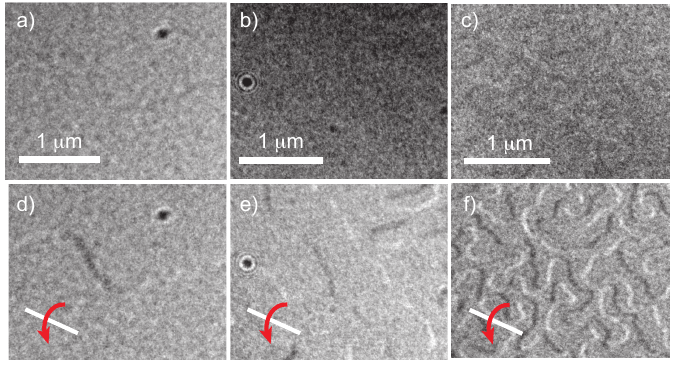}%
\caption{\label{M3} Fresnel mode LTEM images of [Pt/(Co/Ni)$_3$/Ir]$_N$ multi-layers where $N=$ a,d) 2, b,e) 5, and c,f) 10 in the absence and in the presence of a $15^\circ$ sample tilt, respectively.}%
\end{figure}

\noindent Fresnel mode Lorentz TEM images of $M=3$, $N\geq2$ multi-layers show Dzyaloshinskii domain walls (DWs) are formed in these samples; this is indicated by the presence of N\'eel DWs which only display magnetic contrast with the application of a sample tilt (Fig. \ref{M3}). However, DW skyrmions were not observed in thicker films ($N\geq5$) likely due to the presence of hybrid DWs that form due to a large demagnetization field. 

\begin{figure}[h]
\centering
\includegraphics{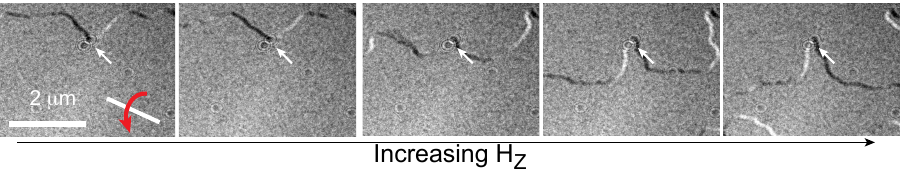}%
\caption{\label{SI_void} Fresnel mode LTEM images of [Pt/(Co/Ni)$_3$/Ir]$_2$ multi-layers in the presence of a perpendicular field applied \textit{in situ}. The domain wall motion is observed to be heavily pinned by the presence of a void (white arrow).}%
\end{figure}

\noindent Additional images depict a void pinning domain wall motion (Fig. \ref{SI_void}). We note that the strong contrast associated with voids was not observed in the case of pinning via domain wall skyrmion. 

